\documentstyle[preprint,aps]{revtex}
\begin{document}
\draft
\preprint{\hfill IFUM -- 94 / 455/ January -- FT}
\title{Passive scalar convection in 2D
 long-range
delta-correlated velocity field: \\Exact results.}
\author
{ Michael Chertkov $^1$,
 Yan V. Fyodorov $^1$ $^2$,
  and Igor Kolokolov $^3$ $^4$}
\address{ $^1$\ Department of Physics of Complex Systems,
Weizmann Institute of Science,
Rehovot 76100, Israel\\
$^2$\ Petersburg Nuclear Physics Institute, 188350 Gatchina,
St.Petersburg, Russia\\
$^3$\ Budker Institute for Nuclear Physics,
Novosibirsk 630 090, Russia\\
$^4$\ INFN, Sezione di Milano, 20133 Milano, via Celoria 16, Italy }
\date{\today}
\maketitle
\begin{abstract}
The letter presents new field-theoretical approach to 2D passive scalar
problem. The Gaussian form of the
distribution  for the Lyapunov exponent is derived and its parameters
are found explicitly.
\end{abstract}
\overfullrule=0pt
\section*{Definition of the model}
The convection of a passive scalar (PS) in 2D by an external long-range
velocity field is well-defined linear problem
\cite{Bat,Kra}. To solve the problem is to find
all correlation functions
of the passive scalar field $w({\bf r};t)$ governed by the
following equation
\begin{equation}
\dot{\omega}+u_{\alpha}\nabla_{\alpha} \omega =\phi ,
\label{i1}
\end{equation}
with $\vec{u}({\bf r};t)$ being an external Eulerian long-range
($q< L^{-1}$) velocity field and $\phi$ being a random external source
localized in $k$ space at $k_0=L^{-1}$. The measure of averaging over
the random source
in the simplest case of white-noise source statistics can be chosen
in the following form
\begin{eqnarray}
& & {\cal D} \phi \exp(-\frac{\int_{-\infty}^{+\infty} \phi^2_{k=1/L}(t) dt}
{2P_2})\ \ \ , \ \ \ \phi_k=(2\pi)^{-2}\int \phi({\bf r};t)
\exp(i{\bf k}{\bf r}) d{\bf r},\nonumber\\
& & <\phi_{{\bf k}}(t)\phi_{ {\bf k}'}(t')>_s\equiv
\Xi_{{\bf k}-{\bf k}'}(t-t')=
\frac{P_2}{2\pi k_0}\delta (t-t')\delta({\bf k}-{\bf k}')
\delta (k-k_0),
\label{i2}
\end{eqnarray}
without any loss of generality as it was shown in \cite{FalLeb}.
Here $P_2$ stands for the flow of the squared PS on the source.

In order to eliminate a homogeneous sweeping it is convenient to pass
to the locally comoving frame \cite{Lvo} expressing the Eq. (\ref{i1})
in terms of the Quasi-Lagrangian (QL) velocities
related to the initial Eulerian
ones by means of the formula
\begin{equation}
{\bf u}({\bf r};t)={\bf v}({\bf r}-\int^t {\bf v}(0;\tau)d\tau;t).
\label{i3}
\end{equation}
The eq.(\ref{i1}) takes correspondingly the form
\begin{equation}
\dot{\omega} +(v^{\alpha}-v_0^{\alpha})\nabla_{\alpha}\omega =\phi,
\ , \
{\bf v}_0={\bf v}(0;\tau).
\label{i4}
\end{equation}
Following the ideas of recent paper \cite{FalLeb}
one can expand $v^{\alpha}({\bf r})-v^{\alpha}(0)=
\sigma^{\alpha\beta} r^{\beta}$ for the points satisfying $r< L$.
Here $\hat{\sigma}$ is the random in time traceless \cite{1*}
 symmetric \cite{2*} matrix. For {\em the white-noise velocity
field statistics} (which is the only case considered
in the present paper) the $\hat{\sigma}$
elements measure has the form
providing space-isotropy of the simultaneous
pair correlator of  the QL velocities
\begin{equation}
 {\cal D} \hat{\sigma}(t)={\cal D}  \left( \begin{array}{cc} a & b \\
b & -a \end{array} \right)=
{\cal D}a {\cal D}b \exp(-S_0), S_0=\frac{1}{2D}
\int(a^2+b^2)dt,
\label{i5}
\end{equation}
 To solve the resulting equation
\begin{equation}
\dot{\omega} +\sigma^{\alpha\beta} r^{\beta}\nabla_{\alpha}\omega =\phi,
\label{i6}
\end{equation}
one performs the substitution
$\omega({\bf r};t)=\psi ({\bf R};t)$, $ R^{\alpha}=W^{\alpha\beta}(t)
r^{\beta}$ that allows to write down
the solution of the eq. (\ref{i6}) in the following form
\begin{equation}
 \omega(\vec{r};t)=\psi({\bf R}(t);t)=\int_{-\infty}^{t}d\tau
\phi (\hat{W}^{-1} (\tau,-\infty)
{\bf R}(t);\tau)\equiv\int_{-\infty}^{t}d\tau
\phi (\hat{W} (t,\tau)\bf{r};\tau),
\label{i8}
\end{equation}
where the matrix $\hat{W}(t;\tau)$ can be sought for as satisfying the
relation
\begin{equation}
\dot{\hat{W}}_t+\hat{W}\hat{\sigma}=0,
\label{i7}
\end{equation}
with initial condition $\hat{W}(\tau,\tau)=\hat{1}$.
Due to Eq.(\ref{i8}) the all order correlation functions of the PS field
 may be rewritten in terms of the correlation functions of the
source field $\phi$. For example, the pair correlation function is
\begin{equation}
 <\omega ({\bf r}_1;t_1)\omega ({\bf r}_2;t_2)>=
\int_{-\infty}^{t_1}d\tau_1
\int_{-\infty}^{t_2}d\tau_2 \Xi (\hat{W}(t_1,\tau_1){\bf r}_1-
\hat{W}(t_2,\tau_2){\bf r}_2;\tau_1-\tau_2)_{\sigma}.
\label{i9}
\end{equation}
The explicit form of the
simultaneous pair correlator ( which has to be isotropic)
 can be extracted from Eqs.(\ref{i9}) and (\ref{i2}) as
\begin{equation}
 <\omega ({\bf r}_1;t)\omega ({\bf r}_2;t)>=
P_2\int_{0}^{\infty}dt
J_0(\frac{\mid\hat{W}(t,0)({\bf r}_1-{\bf r}_2)\mid}{L}),
\label{i10}
\end{equation}
with $J_0(x)$ being the Bessel function of zeroth order.

So, following the calculations of the paper \cite{FalLeb},
we reduced our initial problem to the
evaluation of some functions of the matrix
$\hat{W}$ satisfying the Eq.(\ref{i7}).
The statistics of matrix elements of $\hat{W}(t,0)$ is defined by
the ensemble of random real traceless
symmetric matrices $\hat{\sigma}$, see Eq.(\ref{i5}).
The authors of the paper \cite{FalLeb} have calculated
$<\mid\hat{W}(t,0)r\mid>$ by means of the direct expansion of the
anti-chronological $T-$ exponent that is the formal solution
of the Eq.(\ref{i7}).
It gave them possibility to estimate the simultaneous
pair correlator (\ref{i10}) at small distances $r_{1}-r_{2}$ as
$\frac{P_2}{D}\ln(\frac{L}{r})$.
In addition, analyzing the set of many-point correlation
functions of PS field
Falkovich and Lebedev were able to demonstrate that
the instantaneous statistics of such a field becomes more close to
the gaussian one when passing down-scales.

In the present paper we suggest a new nonperturbative
field-theoretical approach to the
calculation of the PS-field correlations.
We show that the asymptotic behavior of the pair correlator Eq.
(\ref{i10}) is governed
by the self-averaging (or "deterministic") quantity: the Lyapunov
exponent. By means of our method we
rederive, extend and exactly prove all abovementioned results by
Falkovich and Lebedev.
In particular, we find explicitly the value of the variance
of the gaussian distribution characterizing
the short range PS-field statistics.

\section*{Functional representation for averaged functionals of
$\hat{W}$}

The matrix $\hat{W}(t)$ can be extracted from Eq.(\ref{i7}) only
in a form of the anti-chronological time-ordered exponent
\begin{equation}
\hat{W}(t,0)=\tilde{ T}\exp(- \int_{0}^{t}dt'
 \hat{\sigma}(t'))\ , \  \hat{W}(0,0)=\hat{1}.
\label{f1}
\end{equation}
rather than in terms of some regular function of $\hat{\sigma}$.
A similar problem - transformation of a time-ordered exponent of some
linear combination of spin $SU(2)$ operators (arising when one tries
to write down an exact functional
representation for the partition function for
Quantum Heisenberg ferromagnet)
has been solved by Kolokolov \cite{Kol}. The main idea of the
method is to introduce such a new set of integration variables
in the functional integral that $\tilde{ T}\exp$
becomes some regular function when expressed in their terms.
In the present context we use a new modification of the Ansatz
proposed in \cite{Kol} expanding the $\sigma$
matrix over the spin $2\times 2$ matrices as
\begin{equation}
\hat{\sigma}=a\hat{\sigma}_z+b\hat{\sigma}_x.
\label{f2}
\end{equation}

First, we introduce a new basis of the spin algebra:
\begin{equation}
 \hat{\sigma}_y=\left( \begin{array}{cc} 0 & -i \\
i & 0 \end{array} \right)\ , \
\hat{\sigma}_{+}= \hat{\sigma}_z + i\hat{\sigma}_x=
\left( \begin{array}{cc} 1 & i \\
i & -1 \end{array} \right)\ , \
\hat{\sigma}_{-}= \hat{\sigma}_z - i\hat{\sigma}_x=
\left( \begin{array}{cc} 1 & -i \\
-i & -1 \end{array} \right),
\label{f21}
\end{equation}
that corresponds to the rotation of the quantization axis from the usual
orientation (parallel to
the z-axis) to the new one parallel to the y-axis.
Instead of the fields $a(t),b(t)$ we choose
the new ones $\varphi^{\pm}=(a\pm ib)/2$
transforming $\hat{\sigma}$ and the integration measure Eq.(\ref{i5})
to the more compact form:
\begin{equation}
\hat{\sigma}=\varphi^{-}\hat{\sigma}_{+}+\varphi^{+}\hat{\sigma}_{-}
\quad;\quad
 {\cal D} \hat{\sigma}(t)=
{\cal D}\varphi^{\pm} \exp(-S_0)\ , \  S_0=\frac{2}{D}
\int_{0}^{+\infty}\varphi^{+}\varphi^{-}dt,
\label{f5}
\end{equation}

Let us now introduce the operator given in the explicit form as
\begin{eqnarray}
& & \hat{A}(t,0)=
\exp\left[-\hat{\sigma}_-\psi^+(0)\right]
\exp\left[-\hat{\sigma}_{+}\int_{0}^{t}\psi^-(t')
e^{2\int_{0}^{t'} \rho(t'') dt''}dt'\right]
\exp\left[\hat{\sigma}_{y}\int_{0}^{t}
\rho(t') dt'\right]\times\nonumber\\
& &\exp\left[\hat{\sigma}_-\psi^+(t)\right],
\label{f6}
\end{eqnarray}
where $\psi^{\pm},\rho$ are some new dynamical fields.
Using the commutations relations for the spin operators $\sigma_{\pm},
\sigma_{y}$ it is
easy to check that the operator $\hat{A}$ obeys the
differential equation
\begin{equation}
\dot{\hat{A}}_t=\hat{A}
[-\hat{\sigma}_+\psi^-+\hat{\sigma}_-(4\psi^-(\psi^+)^2+
2\rho\psi^++\dot{\psi^+})+\hat{\sigma}_y(-4\psi^-\psi^++\rho)],
\label{f8}
\end{equation}
Using the fact that the first exponential
factor in Eq.(\ref{f6}) makes the operator $\hat{A}$ satisfying
the condition $\hat{A}(0 ,0)=1$ and
comparing the last equation (\ref{f8}) with
the equivalent one (\ref{i7}) we find out the change of variables
\begin{eqnarray}
& & \varphi^-=\psi^-,\nonumber\\
& & \varphi^+=-\dot{\psi^+}+4\psi^-(\psi^+)^2;
\label{f9}\\
& & \rho=4\psi^-\psi^+,
\label{f10}
\end{eqnarray}
in case of which dynamical matrices $\hat{W}$ and $\hat{A}$ coincide.
It allows us to obtain the
explicit functional integral representation for any average written
in terms of $\hat{W}$ by changing variables
from $\varphi^{\pm}$ to $\psi^{\pm}$.
The Jacobian of the map  Eq.(\ref{f9}) is
\begin{equation}
\cal{D}\varphi^{\pm}=\cal{J}[\psi^{\pm}]\cal{D}\psi^{\pm}
,\label{f11}
\end{equation}
and it is essentially dependent
on the kind of regularization and a type of conditions
imposed on the field $\psi^+$. The matter is that
the transformation Eq.(\ref{f9}) contains the derivative of the field
$\psi^+$ with respect to time in its right-hand side.
 Therefore, it should be supplied
 with some initial or boundary conditions. In the papers
\cite{Kol} it has been shown that only imposing conditions of the
initial kind it is possible to ensure the
invertability of the map Eq.(\ref{f9}).
Another point is that in the course of calculation
it is necessary to average some functions of the operator $A(T,0)$ at
fixed time moment $T$ over the measure Eq.(\ref{f5}).
For a given $T$ it is convenient to fix the final
value of the field $\psi^+$
\begin{equation}
\psi^+(T)=-\frac{1}{2}.
\label{boun}
\end{equation}
Then the operator $A(T,0)$ acting on the initial vector
$\left( \begin{array}{c} 1\\0 \end{array}\right)$ produces the following
simple expressions:
\begin{eqnarray}
& &\vec{R}(T)=\hat{A}(T,0) \left( \begin{array}{c} 1 \\ 0 \end{array}\right)=
\frac{1}{2}e^{4\int_{0}^{T}\psi^+\psi^-dt}
\left( \begin{array}{c}
1-2\psi^+(0)
 \\ i(1+2\psi^+(0)
 \end{array}\right),
\label{R}\\
& &
\vec{R}^2(T)=-2\psi^+(0)e^{8\int_{0}^{T}\psi^+\psi^-dt}.
\label{R2}
\end{eqnarray}
Here we exploited the isotropy condition and substituted
$\vec{R}(0)=\left( \begin{array}{c} 1\\0 \end{array}\right)$
for the initial value of the vector $R(T)$ without any loss of
generality.

The regularization of the map Eq.(\ref{f9})
(again as in the papers \cite{Kol,Kol1})
is determined by the physical argumentation. Here it stems from the
translation in time invariance of the white-noise measure Eq.(\ref{f5}).
Indeed, the equality
\begin{equation}
<\varphi^+(T)\int^T\varphi^-(t)dt>=<\varphi^-(T)\int^T\varphi^+(t)dt>
,\label{f12}
\end{equation}
leads to the extension of the definition of the step function
$\theta(x)$ at $x=0$ such that
$\theta(0)=1/2$. Thus, the discrete version of the map (\ref{f5})
($\varphi^{\pm}_n=\varphi^{\pm}(t_n); n=1,...,M ; \epsilon=\frac{T}{M}
\rightarrow 0; t_n=\epsilon n ; M\rightarrow \infty $),
compatible with the
symmetry condition Eq.(\ref{f12}) is
\begin{equation}
\varphi^-_{n}=\psi^-_n \ , \
\varphi^+_n=-\frac{1}{\epsilon}(\psi^+_{n+1}-\psi^+_n)+
\psi^-_n(\psi^+_n+\psi^+_{n+1})^2,
\label{f13}
\end{equation}
that gives the following expression for the Jacobian
\begin{equation}
{\cal J}=const \exp(4\int_{0}^{T} \psi^{+}\psi^{-} dt').
\label{f14}
\end{equation}

When calculating the Jacobian Eq.(\ref{f14}) the fields
$\varphi^{\pm}$ and  $\psi^{\pm}$ were considered to be
independent complex variables
or, in other words, as different coordinate systems in
the whole space ${\cal C}^{2M}$ of the fields' configurations.
The conditions
\begin{equation}\label{surface}
 \varphi^+=(\varphi^-)^*,
\end{equation}
being externally imposed on the model specify
the surface $\Sigma$ in ${\cal C}^{2M}$ along which the
differential forms
${\cal D}\varphi^+\bigwedge {\cal D}\varphi^-$
or ${\cal D}\psi^+\bigwedge {\cal D}\psi^-$
are integrated. For the coordinate sets
$(\psi^{\pm})$ the equation
(\ref{f9}) for $\Sigma$ can be considered to be an implicit one.
According to
the Cauchy-Poincare theorem the integration surface can be deformed
in an arbitrary way in the convergence domain provided we integrate
an analytical function. There exists a continuous family of surfaces
(homotopy)
which includes both the surfaces
$\Sigma$ and the "standard" one ${\Sigma}^\prime$:
\begin{equation}\label{ssurface}
{\Sigma}^\prime=\left\{ -\psi^+=(\psi^-)^*\right\}.
\end{equation}
The explicit expression for such a homotopy
in a more general case can be found in the paper \cite{Kol2}.
Thus, we can replace the surface of integration $\Sigma$
by the standard one ${\Sigma}^\prime$.

Substituting Eq.(\ref{f9}) into the measure Eq.(\ref{i5}) and using
the expressions Eqs.
(\ref{f11},\ref{f14}) we obtain the following modification of the measure
for averaging over $\psi^{\pm}$ \cite{ig1}:
\begin{equation}
 N{\cal D} \psi^{\pm} \exp(-S_1\{\psi^{\pm}\}) \ , \
 S_1=\int_{0}^{+\infty} \left[-\frac{2}{D}\dot{\psi}^+\psi^-+
\frac{8}{D}(\psi^-\psi^+)^2-
4\psi^+\psi^-\right]dt,
\label{w2}
\end{equation}
or , correspondingly, in the discrete form
\begin{equation}
S_1=\sum_n [-\frac{2}{D}(\psi^+_{n+1}-\psi^+_n)\psi^-_n+
\frac{2}{D}\epsilon (\psi^-_n)^2(\psi^+_n+\psi^+_{n+1})^2-
2\epsilon \psi^-_n(\psi^+_n+\psi^+_{n+1})]
.\label{w2d}
\end{equation}
It means that we reformulated the initial problem of the
multiplicative random {\em matrix}
process with the measure Eq.(\ref{i5}) to the multiplicative
random {\em scalar} process with the measure given by Eq.(\ref{w2}).

\section*{Gaussiness of passive scalar correlations}
Another peculiarity of an arbitrary
multiplicative random process is the
Gaussian-like fluctuations of the Lyapunov exponent
$\lambda(T)$
with an amplitude that decays like $const/\sqrt{T}$
when $T$ goes to infinity. Let us note that this fact has been proven
both for the multiplicative random scalar process
(see, e.g. \cite{Fel})  and for matrix one (see
\cite{LeP}.) The usual way to calculate the constant in front of
$1/T^{1/2}$ is to restore
the distribution function of $\lambda(t)$ at arbitrary time
from the set of moments $\langle R^{2n}\rangle$
and then to extract the Gaussian-like peak around
the mean Lyapunov exponent at large enough time
(see for example the calculation
of such a kind in $1D$ \cite{Mel} and quasi $1D$ \cite{FyMi}
localization problem). We follow the same way in the present
context as well.

The Lyapunov exponent in our case is convenient to define as
\begin{equation}
\lambda\equiv \frac{\ln[\vec{R}^2(T)]}{2T}.
\label{Lya}
\end{equation}
In order to find the probability distribution function (PDF) of $\lambda$
let us firstly calculate all the even moments of $\vec{R}$
\begin{equation}
R_m(T)\equiv <R^{2m}>=\frac{<(-2\psi^+(0))^m
e^{8m\int_{0}^{T}\psi^+\psi^-dt}>_1}{<1>_1},
\label{Rm}
\end{equation}
where $<...>_1$ stands for the average with respect to the measure Eq.
(\ref{w2}). Passing to the new integration variables by means of
the gauge transformations
\begin{equation}
\psi^{\pm}=\chi^{\pm}\exp[\mp 4\int_{t}^{T}\chi^+\chi^-dt'
\pm (2m+1)D(T-t)],
\label{ppm}
\end{equation}
that can be written in the discretized form as
\begin{eqnarray}
& &
\psi^+_n=\chi_n^+\exp[-2\sum_{j=n}^M \epsilon (\chi^+_{j+1}+
\chi^+_j)\chi^-_j+(2m+1)\epsilon D (M-n+1)]\nonumber\\
& &\psi^-_n=\chi_n^-\exp[2\sum_{j=n+1}^M \epsilon (\chi^+_{j+1}+
\chi^+_j)\chi^-_j+\epsilon(\chi^+_n+\chi^+_{n+1})\chi^-_n-
(2m+1)\epsilon D (M-n)-\nonumber\\
& &\frac{1}{2}\epsilon(2m+1)D],
\label{ppmd}
\end{eqnarray}
we reduce the averaging in both numerator and denominator (the latter
case corresponds to $m=0$)
of Eq.(\ref{Rm}) to the Gaussian-type ones
\begin{equation}
<(-2\psi^+(0))^m e^{8m\int_0^T\psi^+\psi^-dt}>_1=
(-2)^m<(\chi^+(0))^m>_2 e^{[(2m+2)m+\frac{1}{2}]DT} \ , \
 <1>_1=e^{\frac{DT}{2}},\label{av1}\end{equation}
Here $\langle...\rangle_{2}$ stands for the averaging with respect to the
Gaussian measure
\begin{equation}
 N'{\cal D} \chi^{\pm} \exp(-S_2\{\chi^{\pm}\}) \ , \
 S_2=\frac{2}{D}\int_{0}^{+\infty} \dot{\chi}^+\chi^-dt, \label{w4}
\end{equation}
and one takes into account that the Jacobian
of the map Eq.(\ref{ppm}) is
\begin{equation}
J[\chi]=exp[-2\int_0^T\chi^+\chi^-+(m+\frac{1}{2})D(T-t)].
\label{Jchi}
\end{equation}
The average $<(\chi^+(0))^m>_2$ is equal to $(-\frac{1}{2})^{m}$
due to condition $\chi^+(T)=-\frac{1}{2}$. This result is easy to
get shifting $\chi^+\rightarrow-\frac{1}{2}+\tilde{\chi}^+$ and noticing
that all the averages of $\tilde{\chi}^+$ are equal to zero.
Thus, we arrive at the result
\begin{equation}
R_m(T)=e^{DT(2m+2)m}.
\label{reRm}
\end{equation}

Knowing $R_m$ one can extract Fourier-representation for PDF of $R^2$
in the following way:
\begin{equation}
\tilde{{\cal P}}_k=\sum_{m=0}^{\infty}\frac{(ik)^m}{m!} R_m=
\int_{-\infty}^{+\infty}\frac{e^{-DT/2}}{\sqrt{8\pi DT}}
exp[-\frac{x^2}{8DT}+\frac{x}{2}+ike^x)
\label{PDFk}
\end{equation}
and the PDF for $R^2$ is obtained after the
Fourier inversion in the form
\begin{equation}
\tilde{{\cal P}}(R^2)=
\int_{-\infty}^{+\infty}\frac{dk}{2\pi}e^{-ikz}
\tilde{{\cal P}}_k=\int_{-\infty}^{+\infty}dx
\frac{e^{-DT/2}}{\sqrt{8\pi DT}}exp[-\frac{x^2}{8DT}+\frac{x}{2}]
\delta (R^2-e^x).
 \label{PDFR}
\end{equation}
Correspondingly, the PDF the Lyapunov exponent defined in Eq.(30)
is explicitly given by the expression
\begin{equation}
{\cal P}(\lambda)=\sqrt{\frac{T}{2\pi D}}
e^{-\frac{T(\lambda-D)^2}{2D}},
 \label{PDF}
\end{equation}
that means {\em exact Gaussian fluctuations} around the average value
$<\lambda>=D$ with the variance $\frac{D}{T}$ vanishing
when $T$ tends to infinity. In particular, $\lambda$ tends
asymptotically to a non-random (" deterministic")
quantity in agreement with general theory.

In order to pass from the statistics of
$\lambda$ to the statistics of the
PS-field we introduce the fluctuating quantity
\begin{equation}
Q(\frac{r}{L})\equiv P_2 \int_0^{\infty}dt J_0(\frac{r}{L}e^{\lambda t}),
\label{g10}
\end{equation}
whose distribution can be restored from the PDF of $\lambda$. Indeed,
at $\frac{r}{L}\ll1$ we can cut the integral in the rhs of the Eq.
(\ref{g10}) at $t=\frac{\ln(r/L)}{\lambda}$ and estimate $Q$
with logarithmic accuracy as
\begin{equation}
Q(\frac{r}{L})\approx \frac{P_2}{\lambda(\frac{\ln(L/r)}{\lambda})}
\ln(\frac{L}{r}).
\label{g11}
\end{equation}
It gives us the possibility to extract PDF of $Q$ from the expression
(\ref{PDF}). We obtain
\begin{equation}
{\cal P}(Q)\approx \frac{D}{P_2 \sqrt{2\pi \ln(\frac{L}{r})}}
\exp[-\frac{(Q-\frac{P_2}{D}\ln(\frac{L}{r}))^2}{2P_2Q}D].
\label{g12}
\end{equation}
Such a PDF has at
$r\rightarrow 0$ asymptotically the form of the Gaussian
distribution for the quantity $\frac{Q}{\ln(\frac{L}{r})}$ approaching
the $\delta$-functional form when $\frac{L}{r}\rightarrow\infty$.

\section*{Acknowledgments}
We are grateful to G. Falkovich and V. Lebedev for helpful advises
and numerous discussions. I.K. is grateful to U. Smilansky
for the hospitality extended to him during his stay
in the Weizmann Institute where the main
part of this work was done. Y.V.F acknowledges with thanks the financial
support of the Sir Charles Clore Postdoctoral Fellowship.


\begin{references}
\bibitem{Bat} G.K. Batchelor, J.Fluid Mech. {\bf 5}, 113 (1959)
\bibitem{Kra} R. Kraichnan, Phys. Fluids {\bf 10}, 1417 (1967); J.Fluid
Mech. {\bf 47}, 525 (1971), {\bf 67} ,155 (1975)
\bibitem{FalLeb} G. Falkovich and V.Lebedev, Preprint WIS-93/39/Apr-PH,
Phys.Rev. E (to be published)
\bibitem{Lvo} V. L'vov, Phys. Reports {\bf 207}, 1 (1991)
\bibitem{1*}Due to the fact that one supposes to have an inviscid flow.
\bibitem{2*}It is possible to show that in the case of the white noise
statistics the arbitrary assymetry (which is compatible with the
isotropy of the velocity
field) does not effect all physical averages at all (see also \cite{FalLeb}).
\bibitem{Kol} I. Kolokolov, Ann. of Phys., {\bf 202}, 165 (1990);
I. Kolokolov and E. Podivilov, JETP., {\bf 68},119  (1989)
\bibitem{Kol1}I. Kolokolov, JETP., {\bf 76}, 1099 (1993)
\bibitem{Kol2}I. Kolokolov, Phys. Lett., {\bf 114A}, 99 (1986)
\bibitem{ig1} It should be noted that the convergence is provided
by the term $-\dot{\psi}^+\psi^-$ in the action where the discretization
(\ref{f13}) is assumed.
\bibitem{Fel} W. Feller, An Introduction to probability theory
and its applications (John Wiley and Sons, New York 1957)
\bibitem{LeP} E. Le Page, in Probability Measures on Groups, A. Dold and
B. Eckmann, eds, p.258 (Springer-Verlag 1982)
\bibitem{Mel} V.I. Mel'nikov, JETP Lett, {\bf 32}, 225 (1980)
\bibitem{FyMi} Y.V. Fyodorov and A.D. Mirlin,
 JETP Lett, {\bf 58 }, 615, (1993), 615
\bibitem{Lif} I.M. Lifshitz, S. Gradescul and L.A. Pastur,
Introduction to the theory of Disorded Systems (Wiley, New-York 1988)

\end{references}
\end{document}